\documentstyle[12pt,twoside,fleqn,espcrc1]{article}

\newcommand{\AmS}{{\protect\the\textfont2
  A\kern-.1667em\lower.5ex\hbox{M}\kern-.125emS}}

\title{Transport Coefficients of Quark Gluon Plasma 
       From Lattice Gauge Theory}

\author{S. Sakai\address{Faculty of Education, 
        Yamagata University, Yamagata 990 Japan},\hspace {0.3cm}%
        A. Nakamura\address{Research Institute for Information Science
        and Education, Hiroshima University\\}\hspace{0.2cm}
        and T.Saito$^{\rm a}$}

\begin{document}
\maketitle

\begin{abstract}
   Numerical results for the transport coefficients of
quark gluon plasma are obtained by lattice simulations on 
on $16^3 \times 8$ lattice with the quench
approximation where we apply the gauge action proposed by Iwasaki.
The bulk viscosity is consistent with zero, and the shear viscosity
is slightly smaller than the typical hadron masses. They are not 
far from the simple extrapolation on the figure of perturbative 
calculation in high temperature limit down to $T \sim T_{c}$.
   The gluon propagator in the confined and deconfined phases
are also discussed. 
\end{abstract}

\section{Introduction}

In the phenomenological study of quark gluon plasma(QGP), when its bulk
properties are concerned, the system of quarks and gluons is usually 
treated as gas or liquid.
Then the fundamental parameters of QGP such as transport coefficients,
are very important information. 
In the high temperature limit, they are calculated based on the 
perturbation theory\cite{Kajantie,Horsley,Gavin}. 
However, as temperature $T$ decreases the perturbative calculation
breaks down.  The purpose of this work is to calculate the transport 
coefficient from the fundamental theory 
of QCD by lattice simulations in the vicinity of $T_{c}$.\\
\indent
The calculation of transport 
coefficients on the lattice is formulated in the framework 
of linear response theory
of Kubo\cite{Horsley,Zubarev,Hosoya,Karsch}.
 They are expressed by the space time integral of retarded
Green's function of energy momentum tensors at finite
temperature. The shear viscosity $\eta$
is expressed as,
\begin{equation} \hspace*{1.0cm}
\eta = - \int d^{3}x' \int_{-\infty}^{t} dt_{1} 
e^{\epsilon(t_{1}-t)}
\times \int_{-\infty}^{t_{1}} dt'<T_{12}(\vec{x},t)T_{12}(\vec{x'},t')>_{ret} 
\end{equation}
\indent
Similarly the bulk viscosity and heat 
conductivity are expressed in terms of the retarded Green's 
function of $T_{11}$ and $T_{41}$ components of 
energy momentum tensor.
The direct calculation of the retarded Green's function 
at finite temperature
is very difficult. Then the shortcut is to calculate 
Matsubara Green's function($G_{\beta}$) and   
then by the analytic continuation, we obtain the 
retarded Green's function at finite temperature.
The analytic continuation is carried out by the use of the fact that 
the spectral function of Fourier transform of the both Green's 
functions is the same. 
For the spectral function 
we use the following simplest ansatz\cite{Karsch},
\begin{equation} \hspace*{1.0cm}
\rho(\vec{p}=0,\omega) 
 = \frac{A}{\pi}                                         
(\frac{\gamma}{(m-\omega)^2+\gamma^2}-\frac{\gamma}{(m+\omega)^2+\gamma^2}) 
\end{equation}
\noindent                                                                      
where $\gamma$ is related to the imaginary part of self energy
and partially represents the effects of the interactions.
Under this ansatz, the transport coefficients are calculated as,           
\begin{equation} \hspace*{1.0cm}
\alpha \times a^3 = 2A\frac{2\gamma m}{(\gamma^2+m^2)^2}
\end{equation}  
where $\alpha$ represents $\eta$, $\frac{4}{3} \eta+\zeta$ and
$\chi \cdot T$, and $a$ is lattice spacing.

\indent
Notice that if $m=0$ or $\gamma=0$, transport coefficient                   
becomes zero.
In order to determine these three parameters, at least three
independent data points in $G_{\beta}$
are required in the temperature direction, which means 
$N_{T} \geq 6$.                               

\section{Numerical Results on Transport Coefficients of QGP}
It is found that the the fluctuation of $G_{\beta}$ 
is large, and it is a very important problem to 
reduce the fluctuation of $G_{\beta}$. 
We find that by using the Iwasaki's
improved action, the fluctuation is much reduced\cite{Sakai2}.
Then we apply Iwasaki's Improved action for the 
simulation of SU(3) gauge theory.\\
\indent
 The finite temperature transition point at $N_{T}=8$ with
Iwasaki's improved action is $\beta \sim 2.72$ on $16^3 \times 8$ lattice.
We choose our simulation point at $\beta=3.05, 3.2$ and $3.3$.
From $0.5 \times 10^{6} \sim 10^{6}$ Monte Carlo data, we obtain
$G_{\beta}(t)$ for $T_{11}$ and $T_{12}$ .
But they have still rather large errors. 
The fit of $G_{\beta}$ with parameters
by Eq.(2) is done with SALS.
The shear and bulk viscosities are obtained by these parameters
by Eq.(3), 
and the errors are estimated by the Jackknife method.\\
\indent
The results for shear and bulk viscosities are shown in Fig.1.
It is found that the bulk viscosity is smaller than shear 
viscosity and is consistent with zero within errors.
This is consistent with the result of perturbative calculation at
high temperature limit, which has been $\zeta = 0$\cite{Kajantie,Horsley,Gavin}.\\
\indent
\begin{figure}[htb]
\begin{minipage}[t]{80mm}
\framebox[79mm]{\rule[-26mm]{0mm}{52mm}}
\caption{Shear and Bulk viscosity}
\label{fig:largenenough}
\end{minipage}
\hspace{\fill}
\begin{minipage}[t]{75mm}
\framebox[74mm]{\rule[-26mm]{0mm}{52mm}}
\caption{Shear viscosity from numerical and perturbative calculation}
\label{fig:toosmall}
\end{minipage}
\end{figure}
We compare our results for shear viscosity with those
perturbative formula at high temperature 
limit\cite{Kajantie,Horsley,Gavin},
$ \eta = C \cdot T^{3}/(-\alpha_{s}^{2}log\alpha_{s})$.
Where $\alpha_{s}$ is a running coupling constants given by,
$\alpha_{s}=\frac{2\pi}{11}log(\frac{T}{\Lambda})$,
$\Lambda$ is a scale parameter of QCD.
$C$ is a constant depending on the method of calculation
($0.06 \leq C \leq 0.25 $).
By using the relation $T=1/(N_{T}\times a)$, perturbative formula
is expressed
by the lattice spacing $a$ as follows,
\begin{equation} \hspace*{1.0cm}
\hspace*{2.0cm}\eta a^3 = C /(-N_{T}^{3}\alpha_{s}^{2}log\alpha_{s})
\end{equation}

\indent
   The scale parameter $\Lambda$ on the lattice is determined 
by assuming asymptotic
scaling relation and $\beta_{c}$ at $N_{T}=8$.
From the finite size scaling relation reported by the 
Tsukuba group\cite{Kaneko},  
$\beta_{c} \simeq 2.74$ at $V=\infty$ for $N_{T}=8$. 
And using $T_{c} \simeq 276 MeV$ \cite{Kaneko}
for improved action, $\Lambda$ is determined by, $\Lambda / T_{c} \simeq 1.5 $.
With these quantities the perturbative formula Eq.(4) is compared
with our calculation as shown in Fig.2. It is found that our result are
located not so different from the simple extrapolation of
perturbative calculation on the figure. But notice that the
Eq.(4) breaks down around $T/T_{c} \simeq 1$. \\
\indent
The shear viscosity in the physical unit are obtained if we
assume asymptotic scaling relation for $\beta \geq 2.74$ region.
The results are shown in Fig.3.
They are slightly smaller than the ordinally hadron masses.
What is the physical effects on the phenomenology of quark gluon plasma,
when it has shear viscosity with this magnitudes, is a very 
interesting problem.\\
\indent
In our calculation, $G_{\beta}$ of
$T_{14}$ from which the heat conductivity is calculated, has large 
background and we could not get signal from it.\\

\section{Gluon Propagators at Finite Temperature}

 At first stage of our calculation we planed to calculate transport
coefficients both on confined and deconfined phase.
From the simulation of U(1) gauge theory, it is found that the 
fluctuation of
$G_{\beta}$ in the confined phase is much larger than those of
deconfined phase, and we could not determine the
Green's functions even with about $ 1.5 \times 10^{6}$ data\cite{Sakai}
in confined phase.  
Similar results are obtained in the case of SU(2) gauge theory.
\begin{figure}[htb]
\begin{minipage}[t]{80mm}
\framebox[79mm]{\rule[-26mm]{0mm}{52mm}}
\caption{Shear viscosity in physical unit}
\label{fig:largenenough}
\end{minipage}
\hspace{\fill}
\begin{minipage}[t]{75mm}
\framebox[74mm]{\rule[-26mm]{0mm}{52mm}}
\caption{Gluon propagator in Confined and deconfined phase}
\label{fig:toosmall}
\end{minipage}
\end{figure}
To find the the difference of $G_{\beta}$
between confined and deconfined phase, we study the gluon
propagators at finite temperature on $8^{3} \times 4$ lattice,
because $G_{\beta}$ are 
expressed by the gauge invariant combination of gluon propagators.  
However the gluon propagator itself is
gauge dependent, and we should fix the gauge. It is done by
maximize $I=Re(\sum Tr(U_{\mu}))$ by successive gauge
transformation\cite{Mandula},
which is a lattice version of Lorentz gauge.\\
\indent
The stopping condition for maximizing $I$
is $R=|I^{n+1}-I^{n}|/I^{n} \leq 10^{-14}$.
It is found that
the gluon propagators are very sensitive to $R$ when $R \geq 10^{-9}$.
\indent
With this stopping condition, we could not find the gauge copies
in the deconfined phase. While in the confined phase, gauge 
copies are found. Then in the confined phase, we repeat random
gauge transformation and gauge fixing processes 5 times,  and calculate
the gluon propagators on the gauge fixed configuration with largest 
values of $I$.  In Fig.4, the transverse gluon propagators are
shown for confined and deconfined phase which has a momentum
of $\vec{p}=(2\pi/8,0,0)$ and propagate in the $z$-direction 
which is $8$ in lattice size.\\
\indent
 In the deconfined phase, the gluon propagator is well represented
by the the free Lorentz gauge gluon propagator on the lattice. 
While in the confined phase, the propagator shows
quite different behavior as first found by Nakamura\cite{Nakamura}
at the large distance with standard action.  In Fig.4 it is found that
the effective mass
of gluon seems to increase with the distance. 
What is a physical meaning of this behavior should  further be
investigated, but we find that this is a reason why
the Matsubara Green's function of energy
momentum tensor is so noisy in the confined phase. \\

\noindent
ACKNOWLEDGEMENT  \\
\indent                                      
This work is supported by the Supercomputer Project (No.97-27)
of High Energy Accelerator Research Organization (KEK).       
The simulation is also done on SX-4 at RCNP.
 We would like to express our                                   
thanks to the members of KEK and RCNP for their warm hospitality.

\end{document}